# Experimental set-up for the simultaneous measurement of fission and γ-emission probabilities induced by transfer or inelastic-scattering reactions

R. Pérez Sánchez[1,2], B. Jurado[1*], P. Marini[1,2], M. Aiche[1], S. Czajkowski[1], D. Denis-Petit[1,2], Q. Ducasse[1†], L. Mathieu[1], I. Tsekhanovich[1], A. Henriques[1], V. Méot[2], O. Roig[2]

1) CENBG, CNRS/IN2P3, Université de Bordeaux, 19 Chemin du Solarium, F-33175 Gradignan, France
2) CEA, DAM, DIF, F-91297 Arpajon, France

**Abstract :** Fission and γ-emission probabilities induced by transfer or inelastic scattering reactions with light projectile nuclei are very valuable quantities for constraining the models that describe the de-excitation of heavy nuclei. We have developed an experimental set-up that allows us to simultaneously measure fission and γ-emission probabilities. The measurement of the γ-emission probability at excitation energies where the fission channel is open is challenging due to the intense background of γ rays emitted by the fission fragments. We discuss the procedure to subtract such a background and the constraints that this subtraction and other experimental conditions put on the set up. We show that our set-up complies with these constraints.

## 1. Introduction

Understanding the de-excitation process of heavy nuclei at excitation energies of about 10 MeV is important both for fundamental nuclear physics and for the development of the reaction models used in nuclear astrophysics and in nuclear applications. In this excitation-energy range, an excited heavy nucleus decays via different competing channels: γ-ray emission, particle emission (e.g. neutrons or protons) and fission. The de-excitation process is ruled by fundamental properties of nuclei such as level densities, γ-ray strength functions, particle transmission coefficients and fission barriers. However, when no experimental data are available, the existing nuclear-structure models give very different predictions for these properties, which lead to discrepancies between the calculated cross sections as large as two orders of magnitude or more [1, 2].

Fission probabilities induced by transfer and inelastic-scattering reactions with light projectile nuclei have been used since many years to explore the fission threshold and infer fission-barrier parameters [3, 4]. In particular, transfer- and inelastic-scattering-induced fission probabilities are the only way to obtain information on the fission barriers of fissile nuclei, i.e. nuclei whose fission barrier is lower than the neutron separation energy $S_n$. To constrain the other critical model parameters that rule the competition between fission, γ-ray emission and particle emission, we have developed an experimental set-up that allows one to simultaneously measure fission and γ-emission probabilities. The latter measure the likelihood that the excited nucleus decays only by γ-ray emission. Since the sum of the probabilities of all open decay channels is 100%, the simultaneous measurement of fission and γ-emission probabilities gives also the possibility to constrain the parameters describing particle emission. In addition, the simultaneous measurement of decay probabilities is interesting because the data are collected and analysed under the same conditions, ensuring consistency between them and avoiding the sometimes difficult comparison of results from different experiments.

---

[*] jurado@cenbg.in2p3.fr
[†] Present address: Physikalisch-Technische Bundesanstalt, 38116 Braunschweig, Germany

For measuring γ-emission probabilities, we need to detect the γ rays emitted by the nucleus that is excited by the reaction. This is challenging when the excited nucleus is fissionable because one has to disentangle the γ rays emitted by the decaying nucleus from those emitted in the de-excitation of the fission fragments. The measurement is most difficult for fissile nuclei where the contribution of γ rays emitted by the fission fragments is dominant over a large excitation-energy range.

This work is structured as follows. In section 2 we describe the method used to infer the decay probabilities and identify the requirements that have to be fulfilled by our set-up. In section 3 we present the details of our experimental set-up and show some selected results, which demonstrate its performances in terms of detector response and efficiency. The conclusion and outlook are given in section 4.

## 2. Method to infer the decay probabilities

The decay probability $P$ in the outgoing decay channel $\chi$ (fission or γ-emission) of the excited nucleus $A^*$ produced in the two-body reaction $X(y,b)A^*$ between a projectile $y$ and a target $X$ can be obtained as:

$$P_\chi(E^*) = \frac{N_\chi^C(E^*)}{N^S(E^*) \cdot \varepsilon_\chi(E^*)} \qquad (1)$$

Here, $N^S(E^*)$ is the so-called "singles spectrum", i.e. the total number of detected ejectiles $b$ as a function of the excitation energy $E^*$ of the nucleus $A^*$. $N_\chi^C(E^*)$ is the "coincidence spectrum", corresponding to the number of ejectiles $b$ detected in coincidence with the observable that identifies the decay mode, i.e. a fission fragment or a γ-ray cascade. $\varepsilon_\chi$ is the efficiency for detecting the exit-channel $\chi$ for the reactions in which the outgoing ejectile $b$ is detected. The quantity $N^S(E^*)$ corresponds to the total number of $A^*$ nuclei and $N_\chi^C(E^*)/\varepsilon_\chi$ to the number of $A^*$ nuclei that have decayed via channel $\chi$.

The excitation energy $E^*$ is determined by measuring the emission angle $\theta_{eje}$ and the kinetic energy of the ejectiles $b$, and by applying energy and linear-momentum conservation laws.

The decay probabilities are measured at a given emission angle of the ejectile $\theta_{eje}$. This is interesting because the populated angular momentum of the decaying nucleus $A^*$ depends on this angle. Therefore, the comparison of decay probabilities measured at different ejectile angles can help to study their angular-momentum dependence. This aspect is particularly relevant for studies on the surrogate-reaction method, whose aim is to use decay probabilities induced by transfer or inelastic scattering reactions to infer neutron-induced cross sections of short-lived nuclei that cannot be directly measured [5-9].

## 2.1 Subtraction of γ-ray background

As said above, the γ-emission probability $P_\gamma$ measures the probability that the nucleus $A^*$ decays only through a γ-ray cascade. At excitation energies above $S_n$ or the fission threshold, this probability rapidly decreases with excitation energy $E^*$, because of the competition with the neutron-emission and/or the fission channels. Our aim is to measure the γ-emission probability up to typically $E^*=S_n + 1.5$ MeV. To obtain $P_\gamma$, the measured γ-coincidence spectrum has to be corrected for two backgrounds: (i) the γ rays emitted by the nucleus $(A-1)^*$, which is produced after the emission of one neutron by nucleus $A^*$ and (ii) the γ rays emitted by the fission fragments.

For a given excitation energy $E^*$ of nucleus $A^*$, the γ rays emitted after neutron emission have a maximum possible energy $E^*-S_n$. Therefore, to remove these γ rays we fill the γ-coincidence spectrum with events for which the measured γ-ray energy $E_\gamma^m$ is above a threshold $E_{th}$ equal to the maximum γ-ray energy, i.e. $E_\gamma^m > E_{th}=E^*-S_n$. The γ-cascade detection efficiency $\varepsilon_\gamma^{th}$ is evaluated for each threshold. The validity of this subtraction procedure was demonstrated in [8].

The background of γ rays emitted by the fission fragments is removed by using the expression:

$$N_\gamma^C(E^*) = N_\gamma^{C,tot}(E^*) - \frac{N_{\gamma,f}^C(E^*)}{\varepsilon_f(E^*)} \qquad (2)$$

Here, $N_\gamma^C$ is the final number of γ-ejectile coincidences needed to evaluate the γ-emission probability according to eq. (1), $N_\gamma^{C,tot}$ is the total number of measured γ-ejectile coincidences after application of the $E_{th}$ threshold, $N_{\gamma,f}^C$ is the number of triple fission-fragment/γ-ray/ejectile coincidences and $\varepsilon_f$ is the fission detection efficiency. Thus, $N_{\gamma,f}^C / \varepsilon_f$ is the number of detected γ-rays emitted by the fission fragments. More details on the correction of fission-fragment γ rays can be found in [9]. Determining the triple-coincidences spectrum $N_{\gamma,f}^C$ implies having sufficiently high detection efficiencies, which is challenging. If one wants to limit the number of detectors used, as it is in our case, this requires a very compact geometry.

Let us emphasize that the γ-ray activity of the targets does not generate a background in $N_\gamma^C$ because for the actinides we investigate most of the emitted γ-rays are well below our electronic threshold, which is a few hundred keV, and the few remaining γ-rays can easily be removed by selecting the coincidences with the ejectiles. Reactions on the target support and on light-element target impurities (e.g. oxygen) can, however, generate a background in the γ-coincidence spectrum. We have shown in [8, 9] that we can cope with this background by detecting the ejectiles at backward angles (see section 3.1).

## 2.2 Uncertainties

Our aim is to measure fission probabilities with relative uncertainties $\Delta P_f/P_f < 10\%$ and γ-emission probabilities with $\Delta P_\gamma/P_\gamma < 25\%$. Our previous work has shown that with these uncertainties it is possible to fix model parameters [4] and to investigate the surrogate-reaction method [6,8,9]. In refs. [4] and [9] we have performed a very thorough uncertainty analysis of the decay probabilities, taking into account the covariances between the different measured quantities. Using the equations developed in section G of [9], we can estimate the impact of the uncertainty in the detection efficiencies $\varepsilon_f$ and $\varepsilon_\gamma^{th}$ on the measured decay probabilities. If we make the following assumptions:

-A relative statistical uncertainty of typically 1% in the determination of $N^S$ at a particular emission angle $\theta_{eje}$ and excitation energy $E^*$

-Decay probabilities $P_f = P_\gamma = 50\%$

-A fission efficiency $\varepsilon_f = 60\%$

-A γ-cascade detection efficiency $\varepsilon_\gamma^{th} = 10\%$ at the threshold energy $E_{th}=E^*-S_n$

It results, that we have to determine $\varepsilon_f$ with a relative uncertainty $\Delta\varepsilon_f/\varepsilon_f = 5\text{-}10\%$, and $\varepsilon_\gamma^{th}$ with a relative uncertainty $\Delta\varepsilon_\gamma^{th}/\varepsilon_\gamma^{th}$ of 20%, in order to be able to determine the fission and γ-emission probabilities with a relative uncertainty of 5-10% and 23-25%, respectively. Therefore, we not only need the largest possible fission and γ-cascade detection efficiencies, we also need to know their values with sufficiently good precision.

## 2.3 Main set-up requirements

In the following we summarize the main requirements for the detectors composing our set-up:

-A particle telescope is needed to identify the ejectiles and determine their kinetic energy and emission angle. The latter quantities should be measured with a precision such that the corresponding $E^*$ resolution is of the order of 100 keV. This resolution is needed to investigate the rapid variation of the decay probabilities with increasing $E^*$ near the neutron-emission and the fission thresholds.

-The fission detector should be located as close as possible to the target to maximize the fission efficiency. This implies that the detector should cope with very high rates of elastic scattered beam particles, which can damage the detector and generate pile-up signals that have to be discriminated from those of the fission fragments. In addition, it should have sufficiently good time resolution to make possible the selection of fission events detected in coincidence with the telescope and the γ-ray detectors. Last, it should be segmented in order to allow us to measure the angular distribution of the fission fragments, which has to be taken into account to determine the fission detection efficiency.

-The γ-ray detectors should allow detecting γ rays with energies up to 7-8 MeV (about 1.5 MeV above the $S_n$ of the nuclei we want to investigate) and to discriminate γ rays from neutrons. They should also have sufficiently good timing properties to make possible the selection of coincidence events.

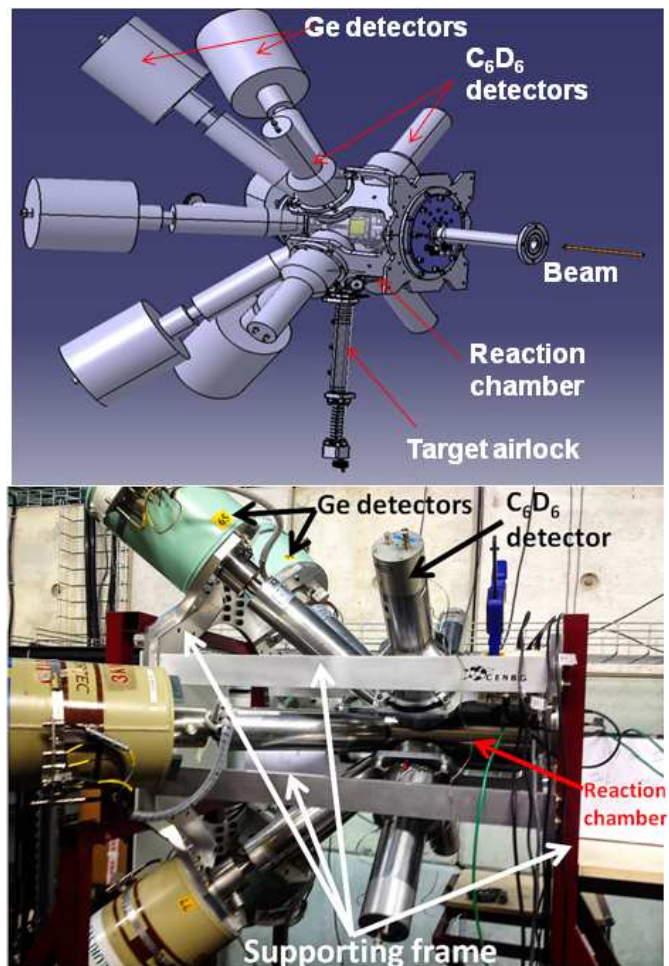

**Figure 1:** (Top) Drawing of the external view of the set-up performed with Computer-Aided Design (CAD) software. (Bottom) Picture of the set-up (side view) during a measurement at the Tandem accelerator of the IPN of Orsay, France. The picture also shows the supporting frame holding the reaction chamber and the γ-ray detectors.

## 3. Experimental set-up

The upper part of figure 1 shows a drawing of the external view of the experimental set-up. A reaction chamber is surrounded by two types of γ-ray detectors: six high-purity germanium detectors and four $C_6D_6$ liquid scintillators. The target airlock, seen in the figure, is needed to isolate the radioactive targets or sources from the environment during the transportation from the target/source handling area to the experimental setup. The airlock is removable and has its own valve. In the handling area, the target is mounted on a target ladder, which is inserted into the airlock. The valve is then closed and the airlock is transported to the experimental area and fixed to the reaction chamber. The lower part of figure 1 shows a picture of the set-up during a measurement at the tandem accelerator of the IPN of Orsay in France.

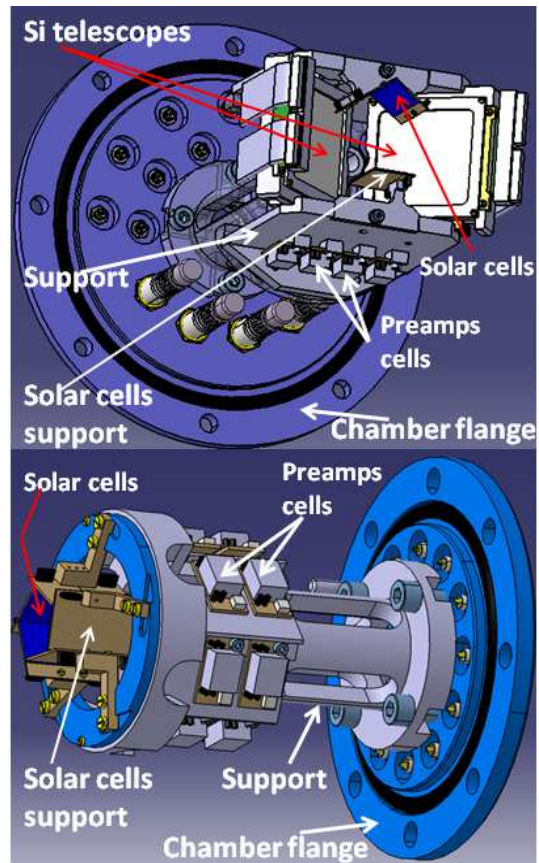

**Figure 2:** CAD drawings of the interior of the vacuum reaction chamber. The detectors located upstream and downstream the target are shown in the upper and the bottom panel, respectively, along with their supports, which are fixed to the chamber flanges.

The reaction chamber is operated at a pressure of about $10^{-6}$ mbar, it houses the target, two position-sensitive particle telescopes ($\Delta E$-$E$) and a fission detector made of solar cells. Our reaction chamber has three openings, two on the sides and one in the center. The central opening is located in the bottom of the chamber and serves to fix the airlock and insert the target ladder. The target ladder allows for three positions. One position is used for the actinide target, which has typical thicknesses ranging from 100 to 200 µg/cm$^2$ and is deposited on a target backing made of natural carbon of 40-100 µg/cm$^2$. The target is placed in such a way that the beam goes first through the carbon backing and then through the target. A second position contains a $^{208}$Pb target of 100 to 200 µg/cm$^2$, which is used for calibration purposes, see below. In the last position we place a target backing alone. Measurements with the target backing are necessary to evaluate and subtract the background coming from reactions taking place on it, see [4, 9]. The airlock includes a high-precision positioning system to place the center of the targets on the beam axis. The inner detectors are placed on either side of the target; they are fixed to the lateral chamber flanges. Figure 2 shows a drawing of the inner detectors together with their supports and the chamber flanges. The two telescopes are located upstream the target. The fission detector is divided in two parts located upstream and downstream the target, each part is composed of several solar cells. In the following sections we give the details of the different detectors composing our set-up.

### 3.1 Particle telescopes

Each telescope is composed of a thin $\Delta E$ detector, whose thickness varies between 100 and 300 µm depending on the experiment, followed by a 5 mm-thick Si(Li) $E$ detector. The $\Delta E$ detectors are 16x16 channels double-sided silicon-strip detectors (DSSSD), with a surface of 5x5 cm$^2$, manufactured by Micron Semiconductor Ltd [10]. As seen in figure 3, we have placed our telescopes at backward angles with respect to the beam direction. This is to minimize the contribution to the singles spectrum from reactions on the carbon backing and on light-element target contaminants. Due to two-body reaction kinematics, ejectiles from these parasitic reactions appear in the excitation-energy spectrum of the nucleus of interest $A^*$ at high excitation energies, thus leaving a large range of $E^*$ in the singles spectrum free of contaminant peaks. The telescopes are centered at 138° (polar angle $\theta_{eje}$) with respect to the beam axis at a distance of 5.5 cm from the target, see figure 3. This was found to be the optimum position considering the available space. The covered polar angles range from 119 to 157° and the associated uncertainty ($\Delta\theta_{eje}$, standard deviation) is about 2°.

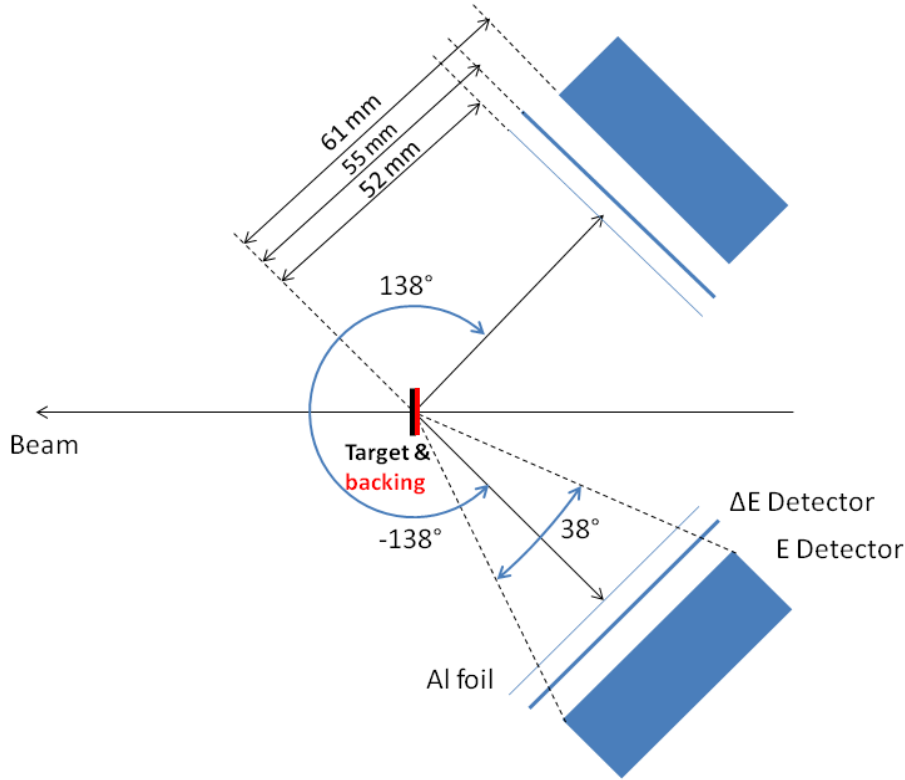

**Figure 3:** Schematic top view of the geometry of the two telescopes.

A 30 µm aluminum foil is placed in front of each $\Delta E$ detector to prevent fission fragments and the alpha particles emitted by the radioactive actinide targets to impinge on the Si detectors. The Al foils are biased to -300 V to repel δ electrons emerging from the target. As shown in figure 3, the $E$ detector has a smaller surface (22.1 cm$^2$) than the $\Delta E$ detector and defines the geometrical efficiency. The sum of the geometrical efficiency of both telescopes is 8.3%. The efficiency of the two telescopes at a given ejectile angle $\theta_{eje} \pm \Delta\theta_{eje}$ is ~ 0.6%, which is sufficient to measure the number of single events with a relative statistical uncertainty of ~1%, as discussed in section 2. Indeed, the cross sections of the reactions that we plan to investigate are of the order of 0.1 mb at backward angles, the beam intensity is about 10-20 enA and the measurement time is typically one week.

Our data acquisition system is triggered when a particle is detected by the $\Delta E$ and $E$ detectors of one of the telescopes. The trigger rate varies from a few tens to a few hundred events per second, depending on the reaction. Hence, the dead-time effects are very small. In any case, the dead time of the acquisition system is the same for the single and the coincidence events. Therefore, it cancels in the ratio of eq. (1) and does not have to be determined.

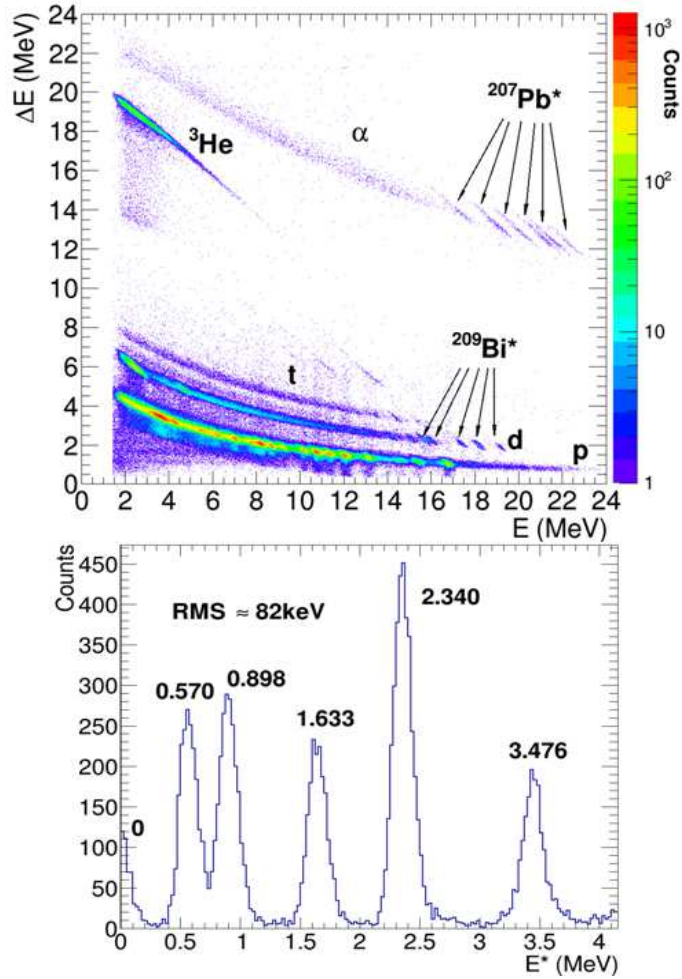

**Figure 4:** (Top) Energy loss $\Delta E$ versus residual energy $E$ of ejectiles measured by one strip of our Si telescopes for the $^3$He + $^{208}$Pb reaction at 24 MeV. The different ejectiles are indicated, the arrows show the states used for the energy calibration of the telescopes. (Bottom) Number of detected alpha particles as a function of the $E^*$ of $^{207}$Pb*, the measured average excitation energy in MeV of each state of $^{207}$Pb* is given together with the root mean square (RMS) deviation.

The measurement of the ejectile kinetic energy is crucial for the determination of the $E^*$ of the excited nucleus $A^*$. In our experiments, each $\Delta E$ strip and the corresponding portion of the $E$ detector are individually calibrated using reactions on $^{208}$Pb. We typically use the $^{208}$Pb target and $^3$He or $^4$He beams to populate the first states of $^{209}$Bi*, $^{207}$Pb* and $^{208}$Pb* via the $^{208}$Pb($^3$He,d)$^{209}$Bi*, $^{208}$Pb($^3$He,$^4$He)$^{207}$Pb* and $^{208}$Pb($^4$He,$^4$He')$^{208}$Pb* reactions, respectively. The $E^*$ of these states is known with high accuracy and the associated kinetic energies of the ejectiles fall in the range of interest for the measurement. The upper part of figure 4 shows a two-dimensional spectrum representing the energy loss $\Delta E$ and the residual energy $E$ of the ejectiles measured by one of the strips of our telescopes for the $^3$He + $^{208}$Pb reaction at 24 MeV incident beam energy. The first states of $^{207}$Pb and $^{209}$Bi used for the calibration are indicated. More details on the calibration procedure can be found in [4]. Similar identification plots as the upper panel of figure 4 are obtained with the actinide targets. However, due to the larger level density of actinides, the first excited states cannot generally be well distinguished.

The lower part of figure 4 shows the number of alpha ejectiles as a function of the $E^*$ of $^{207}$Pb, or in other words, the singles spectrum of $^{207}$Pb*. The root-mean-square (RMS) deviation of the peaks corresponding to the first excited states gives the excitation-energy resolution of our measurements, which ranges between 50 and 100 keV, depending on the experiment. This resolution includes the beam-energy resolution, the energy straggling due to the passage of the ejectiles through different dead layers (target, target backing and Al foil), the variation of the ejectile kinetic energy with $\theta_{eje}$ within the telescope strip, as well as the intrinsic resolution of the $\Delta E$ and $E$ detectors. The obtained excitation-energy resolution is well suited to study the rapid variations of the fission and the γ-emission probabilities at the fission or the particle-emission thresholds.

### 3.2 The fission detector

Solar cells, the devices that are routinely used to convert the energy of sunlight into electricity, represent a particularly interesting option for the detection of fission fragments in our experiments. They were first proposed as fission detectors by Siegert [11], and have been used as fission detectors within large γ-detector arrays as Euroball [12]. The CENBG collaboration has many years of experience using them for the measurement of fission probabilities [4, 6, 13].

The thickness of solar cells varies between 300-500 μm but their depletion depth is less than 1 μm, hence very small, and it cannot be increased by applying a bias voltage because this increases the electronic noise, due to their very low resistivity of a few Ω·cm. The very small depletion depth leads to a very large capacitance, of the order of 40 nF/cm$^2$. To obtain good timing performances with such a high capacitance we use specially designed, current-mode preamplifiers [12]. The typical time resolution that can be obtained is of a few ns, which makes these detectors well suited for coincidence measurements.

Because of the very thin depletion depth, most of the energy deposition occurs in the neutral substrate. The charge collection in the cell is possible thanks to the "funneling" mechanism [14, 15], where the high density of ionization produced along the fragment track locally changes the depletion region into a funnel-like shape extending to the substrate and enclosing the track. This enables the collection of a significant part of the charge produced by the fission fragments via ionization. The funneling efficiency strongly depends on the ionization density profile and it is very small for light particles. Therefore, the response of the cells to light nuclei falls into the detector noise and gives a very impressive pile-up suppression in the fission-fragment region. For this reason, solar cells are much better suited than Si detectors to investigate fission events in the presence of a high background of light charged particles. As mentioned in section 2, this is exactly the situation of our measurements, where the cells are located near the target and are subject to a strong flux of elastic scattered projectile nuclei. Figure 5 shows the spectrum of fission fragments measured by one solar cell in coincidence with the telescopes for the $^{238}$U($^{3}$He,d)$^{239}$Np* reaction. As we can see, the spectrum is not polluted by random coincidences with the elastic scattered beam. Moreover, the double-humped structure due to the different kinetic energies of the light and heavy fission fragments can be clearly distinguished, reflecting the fairly good energy resolution of solar cells. For comparison, figure 4 of ref. [16] shows the fission-fragment energy spectrum measured with a Si detector for the $^{238}$U($^{3}$He,p)$^{240}$Np* reaction at 29 MeV incident energy. The spectrum is polluted by a very intense peak produced by light ions, which clearly overlaps with fission events of low kinetic energies. Also, the separation between heavy and light fission fragments is less pronounced than in our spectrum in figure 5, probably due to the deterioration of the energy resolution caused by radiation damage effects.

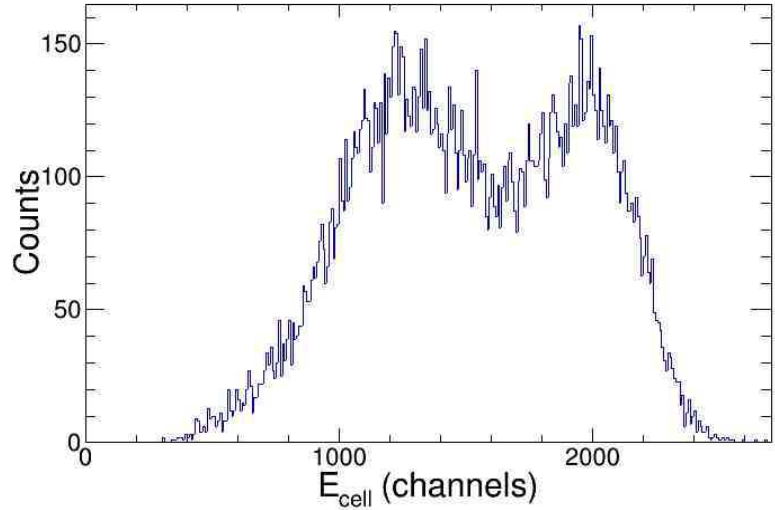

**Figure 5:** Energy spectrum of fission fragments detected in one solar cell in coincidence with deuterons detected in the telescopes for the $^{238}$U($^{3}$He,d)$^{239}$Np* reaction at 24 MeV incident energy.

Indeed, an additional advantage is that solar cells are much more resistant to radiation damage than Si detectors [17, 18]. In [19], we investigated the neutron-induced fission of $^{240}$Pu and $^{242}$Pu target nuclei using solar cells. The cells were subjected during one week to a very intense flux of alpha particles emitted by the Pu target samples. A rate of alpha particles on the cells of 75 kHz only led to a weak performance degradation (see figure 4 in ref [19]). Our aim is to use the present set-up with a $^{240}$Pu target of 100 μg/cm$^{2}$ thickness and 9 mm diameter, corresponding to an alpha activity of 534 kBq. The geometry of our fission detector (see below) is such that the rate of alpha particles on the individual cells will vary between 2.7 and 48 kHz, which is well below the 75 kHz mentioned above. Therefore, no performance degradation is expected from the alpha radioactivity of the $^{240}$Pu target.

Solar cells are also very cost effective and robust. They can be cut into different shapes without showing any deterioration and are thus very well suited to build position-sensitive fission detectors within a very compact geometry. The conducting grid on the surface, which is generally made of very thin silver wires, reduces the sensitive area of the cells because fission fragments are stopped by these wires. The intrinsic efficiency of the cells we use is typically 95%. One can determine the intrinsic efficiency by calculating the ratio of the spontaneous-fission activity of a $^{252}$Cf source measured with the solar cell and with a Passivated Implanted Planar Silicon (PIPS) detector [20], which has no grid and therefore 100% intrinsic efficiency.

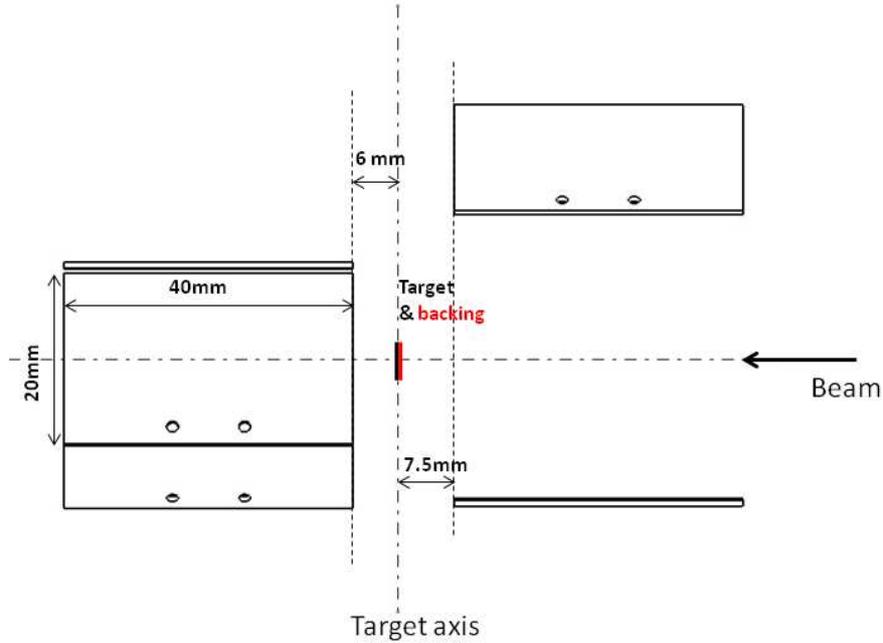

**Figure 6:** Schematic side view of the geometry of the fission detector. Only some of the detector planes are visible.

As shown in figure 6, the fission detector is made of two parts placed upstream and downstream the target. Each part has several detector planes of 40 mm length and 20 mm width each. The downstream part is placed at (6.0 ± 0.4) mm from the target. The distance from the target along the beam axis of the upstream part is (7.5 ± 0.4) mm. We can see in the lower part of figure 7 that the downstream part has the shape of a regular pentagonal prism. The distance from each plane to the beam axis is (12.6 ± 0.3) mm. This distance was chosen to avoid very forward angles where the flux of elastically scattered beam particles is too intense, even for radiation-resistant detectors such as solar cells. Each detector plane is composed of two cells. The length of the two cells composing each side of the prism varies from side to side, i.e. two sides have two cells of (19.5 ± 0.5) mm each, two other sides two cells of (9.5 ± 0.5) and (29.5 ± 0.5) mm, and the remaining side two cells of (29.5 ± 0.5) and (9.5 ± 0.5) mm. With this segmentation we can detect fission fragments at different polar angles $\theta_f$ ranging from 15 to 65°. The upstream part is made of six solar cells located on three planes, see upper panel in figure 7. The distances to the beam axis of the upper and lower detector planes are (28 ± 0.2) and (19.5 ± 0.2) mm, respectively. These distances were chosen to prevent that the cells overshadow the Si telescopes. This geometry leads to an angular coverage $\theta_f$ =110 to 155°. Therefore, with the described segmentation it is possible to measure the fission-fragment angular distribution at forward and backward polar angles. The angular resolution ($\Delta\theta_f$, standard deviation) depends on the size of the cells and varies from 8° for the smallest to 30° for the largest cells.

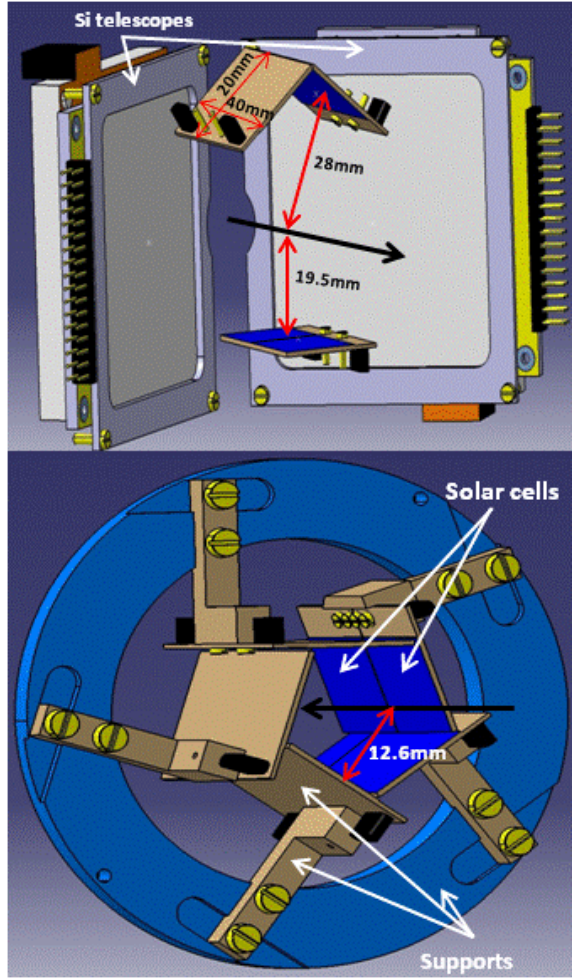

**Figure 7:** Details of the fission detector. The beam is represented by the black arrow. The solar cells are indicated in blue. (Top) Part of the fission detector placed upstream the target. This part is made of three detector planes including six solar cells of various sizes. The Si telescopes are also shown for completeness. (Bottom) Part of the fission detector located downstream the target; it is composed of five detector planes including ten solar cells. The different cell sizes can be observed. The supports of the detector planes are also shown.

### 3.2.1 Fission efficiency

As described above, to obtain the fission and the γ-emission probabilities it is necessary to determine the fission efficiency with good precision. The fission efficiency $\varepsilon_f$ is given by the geometrical efficiency of the fission detector, by the intrinsic efficiency of the solar cells and by the angular anisotropy of the fission fragments. The latter depends on two factors: (i) the angular distribution of the fission fragments in the center of mass reference system, which is determined by the angular-momentum distribution of the fissioning nucleus, and (ii) the kinematical focusing of the fission fragments in the direction of the fissioning nucleus, which depends on the velocity of the fissioning nucleus. We determine the geometrical efficiency with a $^{252}$Cf source of known activity. A Monte-Carlo simulation allows us to include the angular anisotropy effects and infer the "effective" fission efficiency $\varepsilon_f$. The geometrical description of the fission detector included in the simulation is validated with the data obtained with the $^{252}$Cf source. The fission-fragment angular distribution in the center of mass $W(\theta_{f,cm})$, where $\theta_{f,cm}$ is the polar angle of the fragments with respect to the recoiling fissioning nucleus, is obtained by counting the number of fission fragments detected by each cell and dividing this number by the solid angle of the corresponding cell. This solid angle includes the variations due to kinematical effects (e.g. cells that are placed in the direction of the fissioning nucleus detect more fragments) and is determined, together with $\theta_{f,cm}$, via the Monte-Carlo simulation. The experimental angular correlations obtained with our set-up are generally found to be consistent with the form $W(\theta_{f,cm})=a\cdot\cos^2(\theta_{f,cm})+b$, whose parameters $a$ and $b$ are determined by a fit to the measured angular distribution. The coefficients for higher-order terms of $\cos(\theta_{f,cm})$ are very small and do not have a noticeable impact on the quality of the fits. The function $W(\theta_{f,cm})$ is finally inserted into the Monte-Carlo simulation to obtain the effective fission efficiency $\varepsilon_f$. For more details on the determination of the fission efficiency see [4].

We have used our set-up to investigate the $^{240}$Pu($^{4}$He,$^{4}$He')$^{240}$Pu* inelastic scattering reaction at an incident beam energy of 30 MeV. This reaction represents very well the type of reactions we want to study with our set-up. We measured an angular anisotropy of the fission fragments of $W(\theta_{f,cm}=0°)/W(\theta_{f,cm}=90°) = 2.28 \pm 0.32$ at an excitation energy of $^{240}$Pu of 6.5 MeV and a recoil angle of $^{240}$Pu of 19° (corresponding to a kinetic energy of $^{240}$Pu of 1.5 MeV). This anisotropy value is in good agreement with previous results reported in [21]. Our Monte-Carlo simulation shows that when this angular anisotropy is included, the fission efficiency increases from $(55.5\pm4.2)\%$ to an effective fission efficiency $\varepsilon_f = (61.1\pm5.6)\%$. The final uncertainty in the effective fission efficiency is dominated by the uncertainties in the geometry. The obtained efficiency and the relative uncertainty $\Delta\varepsilon_f/\varepsilon_f$ are very close to the reference values given in section 2, implying that with the described set-up it is possible to measure $P_f(\theta_{eje})$ with a relative uncertainty of 10%.

### 3.3 The γ-ray detection systems

As described above, we use two arrays with two different types of γ-ray detectors, four liquid scintillators filled with purified deuterated benzene ($C_6D_6$) and six high-purity germanium detectors. Figure 8 shows the geometrical arrangement of the detectors. The upper panel shows that the central axes of the $C_6D_6$ detectors are included in the plane containing the target. The central axes of the germanium detectors are tilted 45° with respect to this plane. The lower panel shows the angles between adjacent detectors. The distance of the $C_6D_6$ and Ge detectors from the target is 93 and 130 mm, respectively.

One of the main factors that determined our choice for the γ-ray detectors is the presence of neutrons coming from the de-excitation of the heavy decaying nucleus or from the fission fragments. $C_6D_6$ detectors are an alternative to the very usual, non deuterated variety NE-213 liquid scintillators. The replacement of hydrogen by deuterium in the scintillation liquid strongly suppresses γ-rays originating from neutron captures inside the detector itself. Another important advantage of these detectors in the context of our experiments is their ability to provide neutron/$\gamma$-ray discrimination via the pulse-shape analysis technique. Thanks to this, events originating from the interaction of neutrons with the scintillation material can be removed from the measured coincidence spectra. All this makes $C_6D_6$ detectors very well adapted for counting γ-rays with energies up to 7 or 8 MeV in coincidence with the Si telescopes. However, the response function of the $C_6D_6$ detectors is dominated by Compton events and extends from the electronic threshold (in our case about 250 keV) up to the Compton edge, see figure 9. Therefore, $C_6D_6$ detectors are not well suited for accurately measuring the energy of the emitted γ rays.

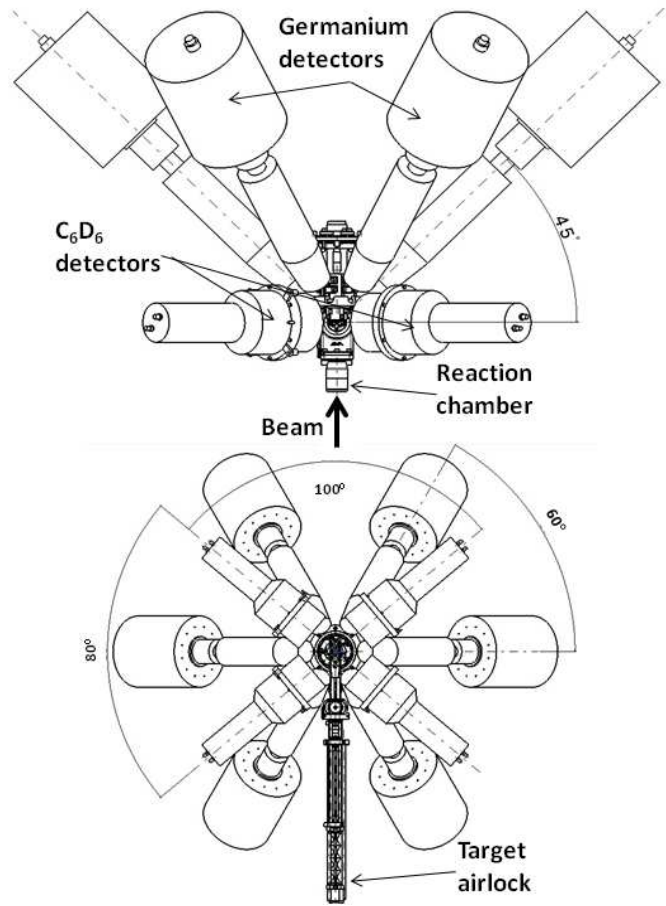

**Figure 8:** Geometrical arrangement of the γ-ray detectors. (Top) Top view of the detector arrays. The angle between the plane containing the central axes of the $C_6D_6$ detectors and the Ge detectors is given. (Bottom) Front view of the detector arrays. The angle between adjacent $C_6D_6$ and adjacent Ge detectors is indicated.

Germanium detectors have a much better energy resolution than $C_6D_6$ detectors (see figure 9). They can be used to measure low-lying γ-ray transition intensities and determine the probabilities for observing specific γ-ray transitions as a function of the excitation energy of the decaying nucleus. As shown recently in [22], this is another valuable observable to constrain models describing the de-excitation process. However, we also want to use the Ge detectors to count γ rays and infer the γ-emission probability. For this reason, we have duplicated the output signals of the Ge detectors. One of the output signals goes to a spectroscopic amplifier with a gain such that the maximum acceptable amplitude of our amplitude-to-digital convertor corresponds to a γ-ray energy of 2 MeV. The other output signal goes to another spectroscopic amplifier with a much lower gain leading to a maximum treatable γ-ray energy of 8 MeV. The high-gain signals are used to study specific transitions, while the low-gain signals are used to infer the γ-emission probabilities. In the case of high-gain signals only the photoelectric component of the measured γ-ray spectrum is considered, whereas in the case of low-gain signals we consider the entire spectrum. It should be noted that pulse-shape discrimination is not possible with Ge detectors. However, for fissile nuclei the neutrons emitted by the fission fragments can be removed by the subtraction of the triple coincidences (see section 2) below $S_n$, provided that the time window selecting the telescope-Ge coincidences is sufficiently large to include the fission neutrons. In this case, the γ-emission probabilities obtained with Ge detectors below $S_n$ are not contaminated by events coming from the interaction of neutrons with the germanium. From the comparison of the telescope-Ge time spectra measured at $E^*$ below and above the onset of fission of $^{240}$Pu*, we deduce that the background of fission neutrons is distributed within a time window of 35 ns and represents less than 14% of the detected events.

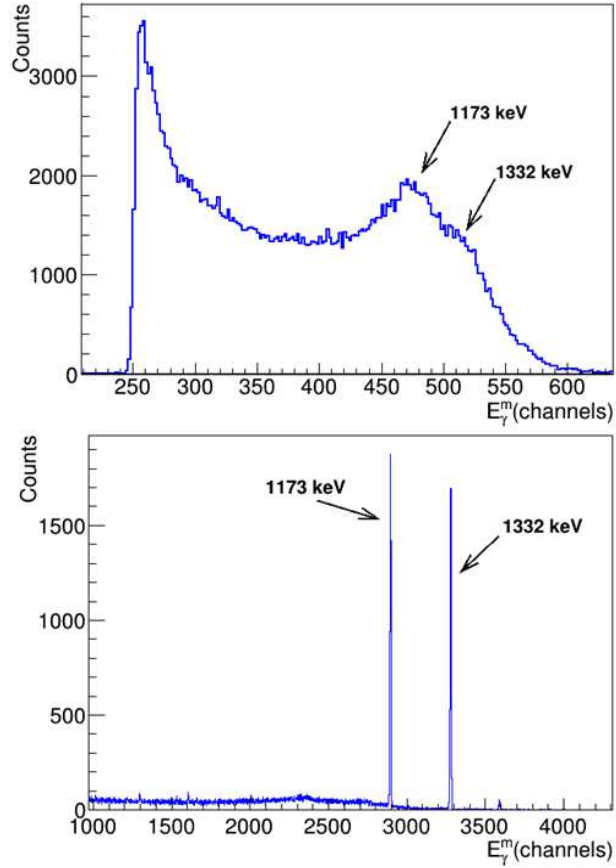

**Figure 9:** Spectra of a $^{60}$Co source measured with a $C_6D_6$ (top) and with a Ge (bottom) detector, the energies of the emitted γ rays are indicated.

### 3.3.1 The γ-cascade detection efficiency

Our γ detector array has a small solid angle and we generally detect only one γ ray of the cascade, in the very few cases where more than one γ ray of the same cascade is detected, we randomly select one of the γ events to fill the coincidence and γ-ray energy spectra. This way, the number of coincidences corresponds to the number of detected γ-ray cascades. To determine the efficiency for detecting a γ-ray cascade, we use the EXtrapolated Efficiency Method (EXEM) that we developed in [23]. In transfer or in inelastic scattering reactions, it is possible to populate excitation energies below the limiting excitation energy, $E^*_{lim}$, where the only possible de-excitation mode is γ-ray emission. Above $E^*_{lim}$, fission and particle emission start to compete with γ-ray emission. Therefore, for excitation energies below $E^*_{lim}$ the γ-emission probability $P_\gamma = 1$ and, according to eq. (1), the γ-cascade detection efficiency, $\varepsilon^{th}_\gamma(E^*)$, can be directly obtained from the ratio between the γ-coincidence and the singles spectra. The EXEM assumes that the measured dependence of the γ-cascade detection efficiency $\varepsilon^{th}_\gamma$ on $E^*$ below $E^*_{lim}$ can be extrapolated to excitation energies above $E^*_{lim}$. We demonstrated the validity of this assumption in [23, 24]. As discussed in section 2.1, a threshold on the γ energy ($E_{th}$) is applied to eliminate γ-rays emitted after neutron emission. The γ-cascade detection efficiency is determined for each $E_{th}$ from the ratio of the corresponding γ-coincidence and singles spectra.

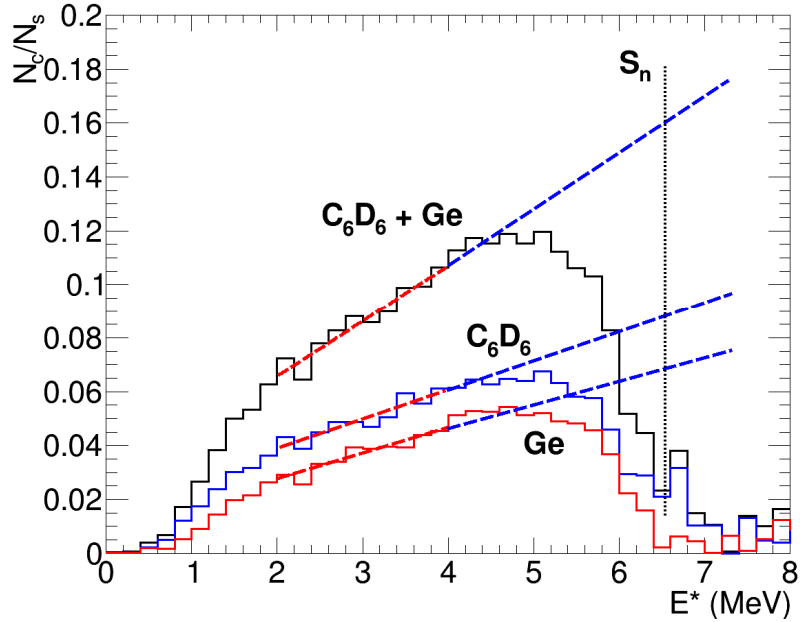

**Figure 10:** Ratio of the γ-coincidence ($E^m_\gamma$>266keV) and the singles spectra as a function of the excitation energy of $^{240}$Pu measured at $\theta_{eje} = 145°$ for the $^{240}$Pu($^4$He,$^4$He')$^{240}$Pu* reaction. The histograms obtained with the coincidences measured by the Ge, the $C_6D_6$ and the sum of all the detectors are represented in red, blue and black, respectively. The vertical dotted line represents the neutron separation energy $S_n$ of $^{240}$Pu. The red dashed lines are linear fit functions adjusted to the data and the blue dashed lines are the extrapolation of the fitted functions to higher excitation energies.

Figure 10 shows the ratio of the γ-coincidence and singles spectra obtained with the $C_6D_6$, the Ge detectors and the sum of all the γ-ray detectors for $E_{th} = 266$ keV. The data were measured for the $^{240}$Pu($^4$He,$^4$He')$^{240}$Pu* inelastic scattering reaction. We can see that all the curves increase with excitation energy up to about 4.5 MeV, above this energy the curves decrease because fission sets in. Indeed, the fission threshold of $^{240}$Pu is significantly lower than its neutron separation energy $S_n$=6.5 MeV. Therefore, to determine $\varepsilon_\gamma^{th}(E^*)$ we fit the data from $E^*$=2 to 4 MeV with a linear function (red dashed lines in figure 10) and take the values of the fit function evaluated at $E^*$ above the fission threshold of 4.5 MeV (blue dashed lines). The uncertainty on $\varepsilon_\gamma^{th}$ is obtained from the uncertainties on the fit parameters. The resulting uncertainty range includes the values of $\varepsilon_\gamma^{th}$ obtained when varying the lower and upper limits of the $E^*$ interval used in the fit by ±400 keV. For the Ge array, the γ-cascade detection efficiency increases from a value of about (4.8 ± 1.4)%, near 4.5 MeV, to about (7.7 ± 2.1)% at $E^*$=7.5 MeV. The efficiency of the $C_6D_6$ array is somewhat larger, varying from nearly (6.0 ± 1.7)% at 4.5 MeV to (9.7 ± 2.3)% at 7.5 MeV. The efficiency of the ensemble of detectors is essentially the sum of the efficiencies of each type of array, it increases from (10.5 ± 2.4)% near 4.5 MeV to (17.9 ± 3.2)% at 7.5 MeV. This feature shows the interest of using also the Ge detectors for counting γ-ray cascades. This will allow us to measure the γ-emission probability between the fission threshold and $S_n$ with efficiencies of the order of 14%.

The efficiencies and the associated relative uncertainties $\Delta\varepsilon_\gamma^{th}/\varepsilon_\gamma^{th}$ obtained with the $C_6D_6$ detectors and with the ensemble of the detectors imply that the γ-emission probabilities can be measured with a relative uncertainty ranging from 24 to 29%. Note that these uncertainties are related to the γ-emission probabilities measured at a particular ejectile angle $P_\gamma(\theta_{eje})$ with the assumptions listed in section 2.2 ($\Delta N_s/N_s$=1% and $P_f=P\gamma$=50%). The expected uncertainties are sufficient to constrain model parameters and investigate the surrogate-reaction method [25]. If the differences between the values of $P_\gamma(\theta_{eje})$ measured at different angles are not significant, the corresponding γ-emission probabilities can be averaged, which will reduce the statistical error and thus the final uncertainty.

## 4. Conclusions and outlook

We have presented a set-up for the simultaneous measurement of fission and γ-emission probabilities induced by transfer and inelastic scattering reactions of light projectile nuclei impinging on heavy target nuclei. Our setup is composed of two position-sensitive Si telescopes to identify the ejectiles and measure their kinetic energies and emission angles. Fission fragments are detected in coincidence with the ejectiles by means of a segmented fission detector made of 16 solar cells. Solar cells are very well adapted for our measurements because they can be arranged in very compact geometries and are rather insensitive to the scattered beam particles. Two arrays of γ-ray detectors allow us to count the number of γ-ray cascades measured in coincidence with the ejectiles. One array is made of four $C_6D_6$ liquid scintillators and the other of six high-purity germanium detectors. Both arrays are used to count γ rays with energies ranging from the electronic threshold to 7-8 MeV. With this set-up we obtain a geometrical efficiency for the Si telescopes of about 8%, a fission efficiency of nearly 61% and a γ-cascade detection efficiency ranging from 6 to 16%, depending on the excitation energy and the detectors used. We have shown that our set-up fulfills the demanding requirements in terms of detection efficiencies necessary to remove the background of γ-rays emitted by the fission fragments and to determine the decay probabilities at various ejectile angles with fairly good precision, about 10% relative uncertainty for the fission probability and between 24 and 29% for the γ-emission probability.

The decay probabilities measured in the experiments carried out with a $^3$He beam incident on a $^{238}$U target, and with $^4$He and $^3$He beams on $^{240}$Pu will be presented in future publications. We also foresee other measurements with the described set-up and $^{242,244}$Pu targets. These future measurements will allow us to provide unique data on the competition between fission and γ emission below the neutron-emission threshold of other fissile nuclei produced by inelastic reactions such as $^{242}$Pu* or $^{244}$Pu*.

**Acknowledgements**

This work was supported by the French "Defi Interdisciplinaire" NEEDS of the CNRS, by the University of Bordeaux and by the European Commission within the 7$^{th}$ Framework Program through CHANDA (Project No 605203).